\documentclass[aps,preprint,showpacs,preprintnumbers,amsmath,amssymb]{revtex4}
\usepackage{amsmath,mathrsfs,amsbsy,color,graphicx,bm,amsthm,amsfonts}
\usepackage{units}
\usepackage{bbm}
\usepackage{times}
\usepackage{dcolumn}
\usepackage{mathrsfs}
\usepackage{amsmath,amssymb,epsfig}
\usepackage{amsmath}
\newcommand{\udots}{\mathinner{\mskip1mu\raise1pt\vbox{\kern7pt\hbox{.}}
\mskip2mu\raise4pt\hbox{.}\mskip2mu\raise7pt\hbox{.}\mskip1mu}}
\begin{document}
\title{Does gravitational wave assist vacuum steering and Bell nonlocality?  }
\author{Shu-Min Wu$^1$\footnote{smwu@lnnu.edu.cn}, Rui-Di Wang$^1$,   Xiao-Li Huang$^1$\footnote{ huangxiaoli1982@foxmail.com }, Zejun Wang$^2$ \footnote{zejunwangcz@foxmail.com (corresponding author)}}
\affiliation{$^1$  Department of Physics, Liaoning Normal University, Dalian 116029, China \\
$^2$  Department of Physics, Changzhi University, Changzhi, 046011, China
}


\begin{abstract}
We study quantum steering and Bell nonlocality harvested by the local interaction of two
Unruh-DeWitt detectors with the vacuum massless scalar field, both in the presence of gravitational waves and in Minkowski spacetime. It is shown  that quantum steerability  under the influence of gravitational waves can be greater than or less than quantum steerability in Minkowski spacetime, which means that the gravitational waves can amplify or degrade the harvested steering. In particular, a resonance effect occurs when the energy gap of the detector is tuned to the frequency of the gravitational wave. We also find that the harvesting-achievable separation range of vacuum steering can be expanded or reduced by the presence of gravitational waves, which depends on the energy gap, the gravitational wave frequency, and the duration of the gravitational wave action.  It is interesting to note that two detector systems that satisfy the Bell inequality in most parameter spaces, regardless of the existence of gravitational waves, indicating that  steering harvesting  cannot be considered to be nonlocal.
\end{abstract}

\vspace*{0.5cm}
\maketitle
\section{Introduction}
Einstein-Podolsky-Rosen (EPR) steering is a quantum phenomenon that describes the control of the state of one entangled subsystem by an observer taking a local measurement of the other, discovered by Schr\"{o}dinger in 1935 \cite{L0-1,L0-2,L1}. Later, Einstein, Podolsky, and Rosen (EPR) studied the concept of quantum steering in their famous paper \cite{L2}, and it is seen as a central concept of the EPR paradox \cite{L3}. Apart from that, Bell established the Bell inequality, which can be used to determine whether quantum nonlocality exists \cite{L4}. Even if two quantum systems are spatially far apart, they cannot be considered nonlocal when the Bell inequality is satisfied. Conversely, if the Bell inequality is violated, then the two quantum systems are said to be nonlocal \cite{L5}. Quantum steerability is a form of medial nonlocal correlation between entanglement and Bell nonlocality \cite{L6}. Entanglement is weaker than quantum steering, and Bell nonlocality is stronger than quantum steering \cite{L7}. Unlike entanglement and Bell nonlocality, quantum steering is characterized by asymmetry \cite{L8}. It has aroused the interest in both theoretically and experimentally, and has received some confirmation \cite{L10,L11,L12,L16,L21,L25}.
It was realized that quantum steering is an important resource for processing quantum information tasks, such as subchannel discrimination, randomness generation, one-sided device-independent quantum key distribution and quantum teleportation \cite{L27,L30,L33,L34}. Like quantum steering, quantum nonlocality is also an important quantum resource, which has been applied to the many fields of communication complexity, quantum cryptography and randomness generation \cite{L35,L36,L37,L38,L39,L40}.

It is well known that the study of quantum information is closely related to the nature of the quantum vacuum. The vacuum state of any quantum field in turn depends on the structure of spacetime. We can obtain understanding of the quantum vacuum and its response to the structure of spacetime by coupling the quantum field to the first-quantized particle detectors. At present, it is shown that the vacuum state of the quantum field has non-local correlations, and these correlations can be harvested using Unruh-DeWitt (UDW) detectors \cite{L60-4,L60-5,L42,L44,L45}. In this way, vacuum correlations can be transferred to the detectors in the form of mutual information, quantum discord, and entanglement. The captured entanglement can be further refined into Bell pairs \cite{L46}, suggesting that the vacuum is in principle a quantum resource for quantum information tasks. This process is called entanglement harvesting \cite{L47,L48,L49,L50,L51,L52,L53,L54,L55,L56,L57,L58,L59,L60,L60-1,L60-2,L60-3}. However, quantum steering and Bell nonlocality from  the vacuum harvesting have not been studied yet in the context of curved spacetime.
Since quantum steering and Bell nonlocality are very important quantum resources, it is necessary to study them from the vacuum harvesting under the influence of gravitational waves. This is one of the motivations for our research. Another motivation is to obtain relevant information of gravitational waves through quantum steering harvesting and Bell nonlocality harvesting.

In this paper, we consider this issue, performing a complete investigation of gravitational waves-assisted steering harvesting phenomenon. In other words, we study quantum steering harvested from the vacuum under the influence of gravitational waves and in Minkowski spacetime. We emphasize that a panoramic understanding of gravitational waves-assisted steering harvesting should include the following two aspects. One is whether steering harvesting can indeed be assisted by gravitational waves in terms of the amount of steering harvested; the other question is whether the achievable separation range for steering harvesting between the detectors can be enlarged in comparison with Minkowski spacetime.  As we will demonstrate, the presence of gravitational waves is actually a mixed blessing for steering harvesting: In some cases, it may contribute to harvesting, while in other cases it may degrade harvesting. Specifically, the presence of gravitational waves can shorten or widen the harvesting-achievable separation
range, depending on the energy gap,  the gravitational wave frequency, and the duration of the gravitational wave action. In addition, we also study vacuum Bell nonlocality  of two UDW detectors and find that Bell nonlocality cannot be extracted from the vacuum in most parameter spaces.

The structure of the paper is as follows. In Sec. II, we briefly introduce quantification of quantum steering and the Clauser-Horne-Shimony-Holt (CHSH) inequality. In Sec. III, we discuss scalar field theory and UDW detectors model in the background of gravitational waves. In Sec. IV, we study vacuum steering and Bell nonlocality of two UDW detectors under the influence of gravitational waves. The last section is devoted to a brief conclusion.
\section{Quantification of quantum steering and CHSH inequality}
Quantum steering represents a distinctive form of quantum correlation. Within a prototypical quantum steering framework, our focus is on two particles that jointly inhabit a bipartite quantum state, named Alice and Bob. It should be noted that the concurrence serves as an effective metric for discerning the quantum entanglement in bipartite states \cite{L61}. Here, we consider the density matrix of the X-state $\rho_{AB}$ as
\begin{eqnarray}\label{S1}
\rho_{AB}=\left(\!\!\begin{array}{cccccccc}
\rho_{11} & 0 & 0 & \rho_{14}\\
0 & \rho_{22} & \rho_{23} & 0\\
0 & \rho_{32} & \rho_{33} & 0\\
\rho_{41} & 0 & 0 & \rho_{44}\\
\end{array}\!\!\right),
\end{eqnarray}
where the real elements of the matrix satisfy $\rho_{ij}=\rho_{ji}$. Then, the concurrence can be expressed as \cite{L63}
\begin{eqnarray}\label{S2}
C(\rho_{AB})=2\max\{|\rho_{14}|-\sqrt{\rho_{22}\rho_{33}}, |\rho_{23}|-\sqrt{\rho_{11}\rho_{44}}\}.
\end{eqnarray}
Subsequently, for the X-state $\rho_{AB}$ shared by Alice and Bob, the steering from Bob to Alice can be sighted if the density matrix $\tau_{AB}$ defined as \cite{L64,L65}
\begin{eqnarray}\label{S3}
\tau_{AB}=\frac{\rho_{AB}}{\sqrt{3}}+\frac{3-\sqrt{3}}{3}(\rho_{A}\otimes\frac{I}{2}),
\end{eqnarray}
is entangled. Here, $\rho_{A}=Tr_{B}(\rho_{AB})$ is the reduced density matrix of Alice, and $I$ represents the two-dimensional identity matrix. Similarly, the corresponding steering from Alice to Bob can be demonstrated if the state $\tau_{BA}$
defined as
\begin{eqnarray}\label{S4}
\tau_{BA}=\frac{\rho_{AB}}{\sqrt{3}}+\frac{3-\sqrt{3}}{3}(\frac{I}{2}\otimes\rho_{B})
\end{eqnarray}
is entangled, where $\rho_{B}=Tr_{A}(\rho_{AB})$ is Bob's reduced density matrix.

Employing Eqs.(\ref{S1}) and (\ref{S3}), the matrix $\tau_{AB}$ of Eq.(\ref{S3}) can be specifically shown  as
\begin{eqnarray}
\tau^{x}_{AB}=\left(\!\!\begin{array}{cccccccc}
\frac{\sqrt{3}}{3}\rho_{11}+m & 0 & 0 & \frac{\sqrt{3}}{3}\rho_{14}\\
0 & \frac{\sqrt{3}}{3}\rho_{22}+m & \frac{\sqrt{3}}{3}\rho_{23} & 0\\
0 & \frac{\sqrt{3}}{3}\rho_{32} & \frac{\sqrt{3}}{3}\rho_{33}+n & 0\\
\frac{\sqrt{3}}{3}\rho_{41} & 0 & 0 & \frac{\sqrt{3}}{3}\rho_{44}+n\\
\end{array}\!\!\right),
\end{eqnarray}
with $m=\frac{(3-\sqrt{3})}{6}(\rho_{11}+\rho_{22})$ and $n=\frac{(3-\sqrt{3})}{6}(\rho_{33}+\rho_{44})$. By using Eq.(\ref{S2}), the state $\tau_{AB}$ is entangled if one of the following inequalities,
\begin{eqnarray}
|\rho_{14}|^{2}>G_{a}-G_{b},
\end{eqnarray}
or
\begin{eqnarray}
|\rho_{23}|^{2}>G_{c}-G_{b},
\end{eqnarray}
is satisfied, where
\begin{gather}
G_{a}=\frac{2-\sqrt{3}}{2}\rho_{11}\rho_{44}+\frac{2+\sqrt{3}}{2}\rho_{22}\rho_{33}+\frac{1}{4}(\rho_{11}+\rho_{44})(\rho_{22}+\rho_{33}),\nonumber\\
G_{b}=\frac{1}{4}(\rho_{11}-\rho_{44})(\rho_{22}-\rho_{33}),\nonumber\\
G_{c}=\frac{2+\sqrt{3}}{2}\rho_{11}\rho_{44}+\frac{2-\sqrt{3}}{2}\rho_{22}\rho_{33}+\frac{1}{4}(\rho_{11}+\rho_{44})(\rho_{22}+\rho_{33}).
\end{gather}
The steering from Alice to Bob can be observed similarly to the steering from Bob to Alice, by examining one of the inequalities,
\begin{eqnarray}
|\rho_{14}|^{2}>G_{a}+G_{b},
\end{eqnarray}
or
\begin{eqnarray}
|\rho_{23}|^{2}>G_{c}+G_{b}.
\end{eqnarray}
According to the above inequalities, we can quantify the steering from Bob to Alice and the steering  from Alice to Bob
\begin{eqnarray}
T^{B\rightarrow A}=\max\{0, \frac{8}{\sqrt{3}}(|\rho_{14}|^{2}-G_{a}+G_{b}), \frac{8}{\sqrt{3}}(|\rho_{23}|^{2}-G_{c}+G_{b})\},
\end{eqnarray}
and
\begin{eqnarray}\label{TAB}
T^{A\rightarrow B}=\max\{0, \frac{8}{\sqrt{3}}(|\rho_{14}|^{2}-G_{a}-G_{b}), \frac{8}{\sqrt{3}}(|\rho_{23}|^{2}-G_{c}-G_{b})\}.
\end{eqnarray}
Note that for a maximally entangled state, such as $|\phi_{AB}\rangle=\frac{1}{\sqrt{2}}(|0\rangle_{A}|0\rangle_{B}+|1\rangle_{A}|1\rangle_{B})$, the normalization factor $\frac{8}{\sqrt{3}}$ is used to ensure that $T^{A\rightarrow B}=T^{B\rightarrow A}=1$.

Bell nonlocality also constitutes a unique manifestation of quantum correlation, which is determined by  Bell inequality. Then, we employ the CHSH inequality, an inequality that holds for two qubits, to test local realist theories. The Bell operator is the key element for the CHSH inequality and is defined as \cite{L66}
\begin{eqnarray}
B_{CHSH}=\mathbf{a}\cdot\mathbf{\sigma}\otimes(\mathbf{b}+\mathbf{b'})\cdot\mathbf{\sigma}+\mathbf{a'}\cdot\mathbf{\sigma}\otimes(\mathbf{b}-\mathbf{b'})\cdot\mathbf{\sigma},
\end{eqnarray}
where $\mathbf{a}, \mathbf{a'} , \mathbf{b}$, and $\mathbf{b'}$ represent unit vectors in $\mathbb{R}^{3}$, and $\mathbf{\sigma} =(\mathbf{\sigma}_{1}, \mathbf{\sigma}_{2}, \mathbf{\sigma}_{3})$ is the vector of Pauli matrices. The expression of the CHSH inequality we mentioned for any bipartite mixed state $\rho$ is
\begin{eqnarray}\label{S5}
B(\rho)=|Tr(\rho B_{CHSH})|\leq2.
\end{eqnarray}
Clearly, the violation of the inequality indicates the presence of Bell nonlocality with respect to the underlying state. For a general two-qubit state, a bound for the maximum possible violation of Eq.(\ref{S5}) is given in  reference \cite{L66-1}.
We need to determine the maximal Bell signal $B(\rho)$, which in the case of two-qubit systems $\rho$, can be considered equivalent to
\begin{eqnarray}\label{S6}
B(\rho)=2\sqrt{\max_{i<j}(F_{i}+F_{j})},
\end{eqnarray}
where $F_{i}$ and $F_{j}$ are the two largest eigenvalues of $U(\rho)=T^{T}_{\rho}T_{\rho}$, and $T=(t_{ij})$ denotes the  correlation matrix with $t_{ij}=Tr[\rho\sigma_{i}\otimes\sigma_{j}]$. For two-qubit X-state, the corresponding matrix $U(\rho)$ has three eigenvalues
\begin{gather}\label{S6-1}
F_{1}=4(|\rho_{14}|+|\rho_{23}|)^{2},\nonumber\\
F_{2}=4(|\rho_{14}|-|\rho_{23}|)^{2},\nonumber\\
F_{3}=(|\rho_{11}|-|\rho_{22}|-|\rho_{33}|+|\rho_{44}|)^{2}.
\end{gather}
Since $F_{1}$ is larger than $F_{2}$, the maximal Bell signal reads
\begin{eqnarray}\label{S7}
B(\rho_{x})=\max\{B_{1},B_{2}\},
\end{eqnarray}
with $B_{1}=2\sqrt{F_{1}+F_{2}}$ and $B_{2}=2\sqrt{F_{1}+F_{3}}$ \cite{L67,L68}.
Therefore, we will use Eqs.(\ref{S5}), (\ref{S6}), and (\ref{S7}) to judge whether quantum steering is nonlocal in the context of gravitational waves.

\section{Unruh-DeWitt model in the background of gravitational waves}
In the context of a gravitational wave propagating along the z-direction, the pertinent description is encapsulated within the line element as follows:
\begin{eqnarray}\label{S8}
ds^2 &=& -dt^2 +dz^2 +(1+A \cos\left[\omega (t-z)\right]) dx^2  + (1-A \cos\left[ \omega(t-z)\right])dy^2 \nonumber\\
&&=- du dv +(1+ A \cos \omega u ) dx^2  + (1-A \cos \omega u)dy^2.
\end{eqnarray}
Here, we introduce light cone coordinates, denoted as $u:=t-z$ and $v:=t+z$, defined based on Minkowski coordinates $(t, x, y, z)$. This spacetime solution corresponds to the linearized field equation and is effective for the leading order of  $A\ll1$. Within this spacetime framework, we examine a massless scalar field $\Phi(x)$ that satisfies to the Klein-Gordon equation at a given spacetime point
\begin{eqnarray}\label{S9}
\Box \phi(\mathsf{x}) = 0,
\end{eqnarray}
where the d'Alembertian operator $\Box$ is related to the Eq.(\ref{S8}). Solving this equation in light-cone coordinates
$x=(u, v, x, y)$ produces a complete set of solutions \cite{L69}
\begin{eqnarray}\label{S10}
u_{\vec{k}}(\mathsf{x}) =& \frac{\gamma^{-1}(u )}{\sqrt{2k_{-}}(2\pi)^{\frac{3}{2}}} e^{ik_a x^a -ik_{-}v  -\frac{i}{4k_{-}}\int_0^u du \,  (g^{ab}k_a k_b )},
\end{eqnarray}
where the constants $k_{-}$ and $k_{a}$ emerge as separability constants upon solving Eq.(\ref{S10}) in the light-cone coordinates, the indices $a$ and $b$ run over  $\{x, y\}$, and $\gamma^{-1}(u) = [\det g_{ab}(u)]^{\frac{1}{4}}$. We interpret the mode functions in Eq.(\ref{S10}) as delineating the perturbation to the Minkowski vacuum caused by the gravitational wave.

Through detailed calculations, we can obtain the vacuum Wightman function as
\begin{eqnarray}
W(\mathsf{x},\mathsf{x}')&:=& \langle 0|\phi(\mathsf{x}) \phi(\mathsf{x}') | 0\rangle = \int d \mathsf{k} \, u_{\mathsf{k}}(\mathsf{x}) u_{\mathsf{k}}^*(\mathsf{x}')= W_{ \mathcal{M}}(\mathsf{x},\mathsf{x}') + W_{\rm GW}(\mathsf{x},\mathsf{x}').
\end{eqnarray}
Here, $W_{ \mathcal{M}}(\mathsf{x},\mathsf{x}')$ denotes the Minkowski  Wightman function, independent of the gravitational wave, in light-cone coordinates, and $W_{\rm GW}(\mathsf{x},\mathsf{x}')$ represents the first-order modification of the Minkowski Wightman function with respect to the amplitude of the gravitational wave  \cite{L70}. Their specific expressions, respectively, are
\begin{eqnarray}
W_{\mathcal{M}} (\mathsf{x},\mathsf{x}') &=& \! \frac{1}{4\pi i \Delta u  }    \delta\left(\frac{\sigma_{\mathcal{M}}(\mathsf{x},\mathsf{x}')}{\Delta u}  \right)
+ \frac{1}{4\pi^2  \sigma_{\mathcal{M}}(\mathsf{x},\mathsf{x}')} ,
\end{eqnarray}
and
\begin{eqnarray}
W_{\rm GW}(\mathsf{x},\mathsf{x}') &=&\! -\tfrac{A}{4 \pi^{2}  }  \rm{sinc} \left( \tfrac{ \omega}{2} \Delta u  \right)  \cos  \left( \tfrac{\omega}{2}    [u+u'] \right) \times\! \tfrac{ \Delta x^2 - \Delta y ^2  }{\Delta u^2}\!
 \left[ i \pi \delta'\!\left(  \tfrac{\sigma_{\rm \mathcal{M}}(\mathsf{x},\mathsf{x}')}{\Delta u} \right)  \! + \!\tfrac{\Delta u ^2}{\sigma_{\mathcal{M}}^2(\mathsf{x},\mathsf{x}') }  \right]\!,
\end{eqnarray}
where  $\Delta \mathsf{x}^\mu := \mathsf{x}^\mu-\mathsf{x}'^\mu$, and $\sigma_{\mathcal{M}}(\mathsf{x},\mathsf{x}') := - \Delta u\Delta v+  \Delta x^2  +  \Delta y^2$ represents the geodesic distance between points $x$ and $x'$ in Minkowski space.

To experimentally investigate the impact of the gravitational wave on vacuum steering and Bell nonlocality of a scalar field theory, we utilize two Unruh-DeWitt detectors, labeled $A$ and $B$, with the ground state $|0_{D}\rangle$ and the excited state $|1_{D}\rangle(D\in\{A,B\})$, separated by the energy gap 2$\Omega$. Two detectors interact locally with the scalar field $\phi(\mathsf{x})$ along its trajectory $\mathsf{x}_{D}(t)$ of parametrized by its proper time $t$. The Hamiltonian governing this interaction is
\begin{eqnarray}\label{S11}
H_D(t) &= \lambda \chi\! \left(t \right)\Big(e^{ i\Omega t} \sigma^+  +  e^{- i\Omega t}\sigma^- \Big) \otimes  \phi\left[\mathsf{x}_D(t)\right],
\end{eqnarray}
where $\chi(t) = \exp[-\frac{(t-t_0)^2}{2\sigma^2}]$ is the switching function of $t_{0}$ and $\sigma$, which correspond to the occurrence and duration of the interaction, respectively, $\lambda$ is the coupling constant and describes the strength of the interaction, and $\sigma^+ =|{1_D}\rangle\langle{0_D}|$ and $\sigma^-=|{0_D}\rangle\langle{1_D}|$ denote SU(2) ladder operators  acting on the detector Hilbert space.

Initially, consider these detectors prepared as  ($t\rightarrow-\infty$) in their ground state $|{0} \rangle_A|{0} \rangle_B$, and the field state in a suitably defined the vacuum state $|0\rangle$, ensuring that the combined state of the detectors and the field is denoted as $|{\Psi_i}\rangle = |{0} \rangle_A|{0} \rangle_B|{0}\rangle$. With the interaction Hamiltonian of Eq.(\ref{S11}), the final state ($t\rightarrow\infty$) of the detectors and field becomes $|{\Psi_f}\rangle = \mathcal{T} e^{  -i \int_{\mathbb{R}}  dt\, \left[ H_A( t) +  H_B(t) \right] } |{\Psi_i}\rangle$, where $H_A( t)$ and $H_B(t)$ are specified in Eq.(\ref{S11}), and $\mathcal{T}$ represents the time ordering operator. The reduced state of the detectors is derived in the basis $\{ |{0_A 0_B}\rangle, |{0_A 1_B}\rangle, |{1_A 0_B}\rangle, |{1_A 1_B}\rangle \}$ by performing a trace operation over the field
\begin{eqnarray}\label{AB}
\rho_{AB} :=\rm{tr}_{\phi}(|\Psi_{f}\rangle\langle\Psi_{f}| )
= \begin{pmatrix}
1 - 2 P  & 0 & 0 & X \\
0 & P  & C & 0 \\
0 & C^* & P & 0 \\
X^* & 0 & 0 & 0
\end{pmatrix} + \mathcal{O}\!\left(\lambda^4\right).
\end{eqnarray}
Here, the matrix elements $X$ and $C$ characterize nonclassical correlations and are the sum of two terms: $X= X_{\mathcal{M}} + X_{\rm GW}$ and $C = C_{\mathcal{M}} + C_{\rm GW}$. Obviously, $X_{\mathcal{M}}$ and $C_{\mathcal{M}}$ correspond to the values of $X$ and $C$ when the detector is in Minkowski space, $X_{\rm GW}$ and $C_{\rm GW}$ correspond to the modifications of the matrix elements $X$ and $C$ resulting from the influence of the gravitational wave. Besides, there is also a matrix element, the transition probability $P$ that in the distant future ($t\rightarrow\infty$) the detector will transition to the excited state $|1_{D}\rangle$, as a result of the vacuum fluctuations and a limited interaction time.

In particular, consider the scenario where the detector is at rest relative to the Minkowski coordinates introduced in Eq.(\ref{S8}), resulting in its trajectory being a geodesic $\mathsf{x}_D(t)=(t, 0, 0, 0)$. Due to the condition $\Delta x^2= \Delta y^2=0$, the transition probability $P$ is not affected by gravitational waves. It can also be asserted that a single detector is unable to discern the existence of a gravitational wave. Thus, we can specifically express $P$ as
\begin{eqnarray}
P = \frac{\lambda^2}{4 \pi} \left[ e^{-\sigma^2 \Omega^2} - \sqrt{\pi} \sigma \Omega \left(1- \rm{erf}[\sigma \Omega]\right)\right].
\end{eqnarray}
To operationally probe vacuum steering and Bell nonlocality across spacetime regions, we consider the trajectories of detectors $A$ and $B$, delineated in Minkowski coordinates $\mathsf{x}_A(t)=(t, 0, 0, 0), \mathsf{x}_B(t)=(t, D, 0, 0)$. Consider that due to the detectors interacting with the field for an approximate proper time $\sigma$, detectors following these trajectories may be deemed approximately spacelike separated during the interaction when $D > \sigma$ and timelike when $D < \sigma$; $D$ denotes the average appropriate distance between the detectors  \cite{L70}. Combined with the Wightman function, the $X_{\mathcal{M}}$, $C_{\mathcal{M}}$, $X_{\rm GW}$, and $C_{\rm GW}$ can be expressed as
\begin{eqnarray}
X_{\mathcal{M}} := i \frac{\sigma \lambda^2}{4D \sqrt{\pi}} e^{-\sigma^{2}\Omega^2 -2i\Omega t_{0}- \frac{D^{2}}{ 4\sigma^{2}}} \left[ \rm{erf} \left(  \frac{iD}{2\sigma}\right) -1 \right]\!,
\end{eqnarray}
\begin{eqnarray}
C_{\mathcal{M}} := \frac{\sigma \lambda^2}{4 D\sqrt{\pi}} e^{-\frac{D^{2}}{ 4\sigma^{2}} }\times \left(\rm{Im} \left[ e^{i D \Omega} \rm{erf} \left( i \frac{D}{2\sigma} +  \sigma \Omega \right)\right]   - \sin  \Omega D \right),
\end{eqnarray}
\begin{eqnarray}
X_{\rm{GW}}:= \frac{ A \sigma \lambda^2 }{ 4D^2\pi^{3/2}}  f(\omega, \Omega, \sigma,t_0)   \left(I_1 + I_2 \right),
\end{eqnarray}
and
\begin{eqnarray}
C_{\rm{GW}}:= -\frac{A \sigma \lambda^2  }{ 4D^2\pi^{3/2}}  e^{- \frac{ \sigma^{2}\omega^{2}}{4}}  \cos \left(\omega t_{0}\right)  \left( I_3 + I_4 \right),
\end{eqnarray}
with
\begin{eqnarray}
f(\omega, \Omega, \sigma,t_0):=e^{-\frac{\sigma^2}{4}(\omega - 2\Omega)^2 - it_0(\omega+ 2\Omega)}+e^{-\frac{\sigma^2}{4}(\omega +2\Omega)^2 + it_0 ( \omega- 2\Omega )}.
\end{eqnarray}
The terms $I_{1}$ and $I_{2}$ are about $\omega$, $D$, and $\sigma$ of complex functions, $I_{3}$ and $I_{4}$ are cumbersome functions of $\omega$, $D$, $\sigma$, and $\Omega$. Their representations are as follows:
\begin{gather}
I_{1}=i \frac{ \pi e^{-\frac{ D^2 }{ 4\sigma^{2}}}}{ \omega} \left[\left(\frac{D^{2}}{4\sigma^{2}}+1 \right)\sin\left( \tfrac{ \omega}{2} D\right)- \frac{D  \omega}{4} \cos \left( \tfrac{ \omega}{2} D\right)     \right] ,\nonumber\\
I_{2}=\frac{\pi }{ \omega} \left(   \rm{erf} \left(\frac{\sigma  \omega }{2}\right) -
 e^{-\frac{D^2}{4\sigma^2}}  \rm{Re} \left[e^{i \frac{\omega}{2} D } \left(1 + \frac{D^2}{4 \sigma^2}-i \frac{D  \omega}{4}  \right) \rm{erf} \left( \frac{\omega}{2} \sigma +  \frac{ i \rm{D} }{2 \sigma } \right) \right] \right),\nonumber\\
I_{3}=\frac{ \pi e^{- \frac{D^2}{4 \sigma^2}}}{4  \omega } \Bigg[\left(D \omega +2 D \Omega\right) \sin \left(D\left[\frac{\omega}{2}+\Omega\right]\right) +  \left(D \omega -2 D \Omega\right) \sin \left(D\left[\frac{\omega}{2}-\Omega\right]\right) \nonumber\\
+ \left(\frac{D^{2}}{\sigma^{2}}+4\right)\left( \cos \left(D \left[ \frac{\omega}{2} + \Omega\right]\right) - \cos \left(D \left[\frac{\omega}{2} - \Omega\right]\right)\right)\Bigg],\nonumber\\
I_{4}=\frac{\pi}{ \omega}\left( \rm{erf} \left[\sigma \left( \frac{ \omega}{2}-\Omega \right) \right]+\rm{erf}\left[\sigma\left( \frac{ \omega}{2}+\Omega \right)\right]-  e^{-\frac{D^{2}}{4\sigma^{2}}}\rm{Re}\left[ Q_{+}R_{+} +  Q_{-}R_{-} \right] \right) \nonumber,
\end{gather}
with $ Q_{\pm}:=-i e^{i D \left(\frac{\omega}{2} \pm \Omega \right)}\rm{erf}{\left[i\frac{D}{2\sigma} + \sigma \left( \frac{\omega}{2} \pm \Omega \right)\right]}$ and $R_{\pm}:=\frac{D}{2} \left( \frac{\omega}{2} \pm  \Omega\right) + i \left(1 + \frac{D^2}{4\sigma^{2}} \right) $.

\section{Quantum steering and Bell nonlocality of two Unruh-DeWitt detectors}
In this section, we will consider how gravitational waves affect vacuum steering and Bell nonlocality, exploring gravitational wave-assisted steering harvesting and the harvesting-achievable range of gravitational waves extending quantum steering. Employing Eqs.(\ref{TAB}) and  (\ref{AB}), we can obtain an analytic expression of the steering $T^{A\rightarrow B}$ as
\begin{equation}
T^{A\rightarrow B}=\max \{0, \frac{8}{\sqrt{3}}(|X|^{2}-\frac{\sqrt{3}}{2}P^2-\frac{1}{2}P), \frac{8}{\sqrt{3}}(|C|^{2}+\frac{\sqrt{3}}{2}P^2-\frac{1}{2}P)\}.
\end{equation}
Here, the expressions for both quantum steering, $T^{B\rightarrow A}$ and $T^{A\rightarrow B}$, are identical. Therefore, our analysis will focus solely on the steering $T^{A\rightarrow B}$.
To observe the direct impact of gravitational waves on vacuum steering, we also investigate the steering $S^{A\rightarrow B}$ in Minkowski spacetime.

\begin{figure}
\begin{minipage}[t]{0.5\linewidth}
\centering
\includegraphics[width=3.0in,height=5.2cm]{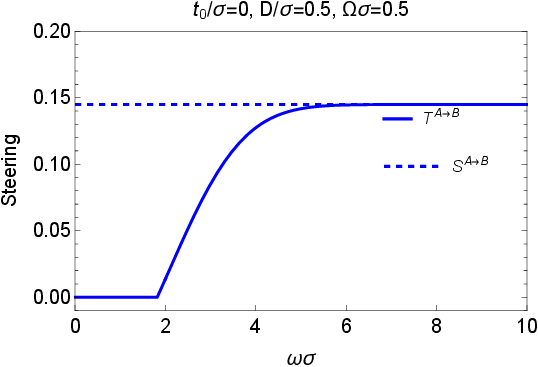}
\label{fig1a}
\end{minipage}%
\begin{minipage}[t]{0.5\linewidth}
\centering
\includegraphics[width=3.0in,height=5.2cm]{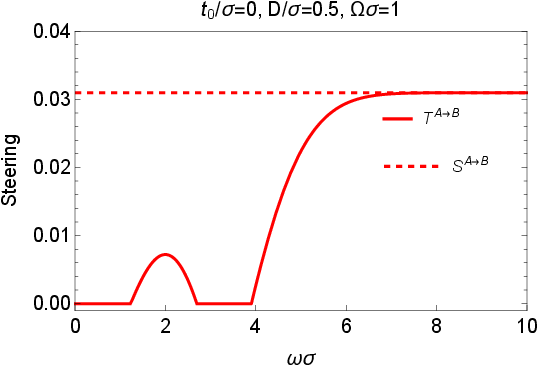}
\label{fig1b}
\end{minipage}%

\begin{minipage}[t]{0.5\linewidth}
\centering
\includegraphics[width=3.0in,height=5.2cm]{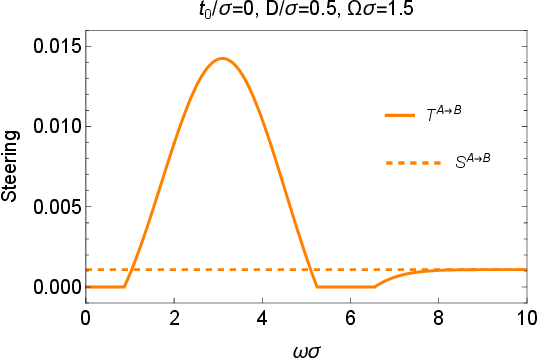}
\label{fig1c}
\end{minipage}%
\begin{minipage}[t]{0.5\linewidth}
\centering
\includegraphics[width=3.0in,height=5.2cm]{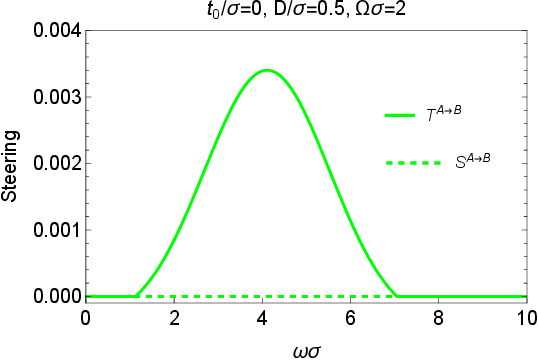}
\label{fig1d}
\end{minipage}%

\begin{minipage}[t]{0.5\linewidth}
\centering
\includegraphics[width=3.0in,height=5.2cm]{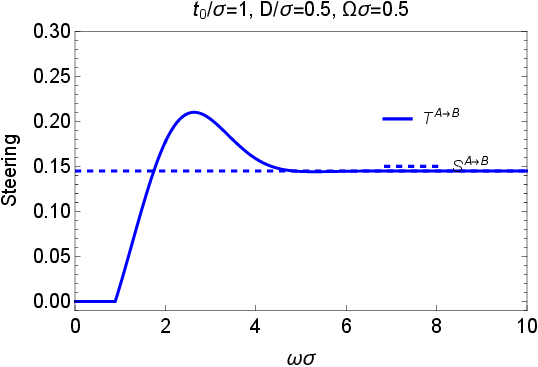}
\label{fig2a}
\end{minipage}%
\begin{minipage}[t]{0.5\linewidth}
\centering
\includegraphics[width=3.0in,height=5.2cm]{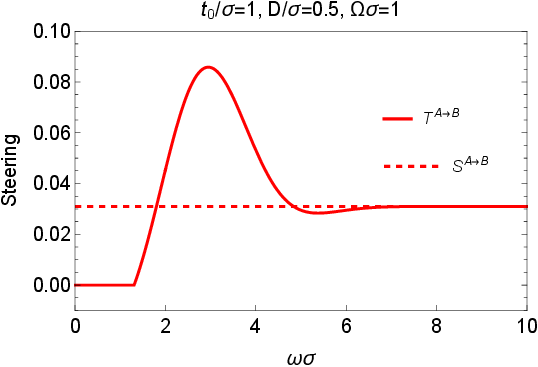}
\label{fig2b}
\end{minipage}%

\begin{minipage}[t]{0.5\linewidth}
\centering
\includegraphics[width=3.0in,height=5.2cm]{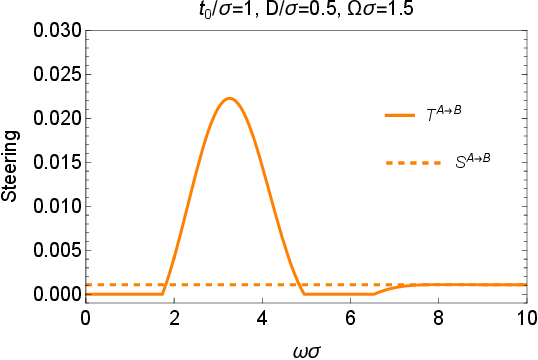}
\label{fig2c}
\end{minipage}%
\begin{minipage}[t]{0.5\linewidth}
\centering
\includegraphics[width=3.0in,height=5.2cm]{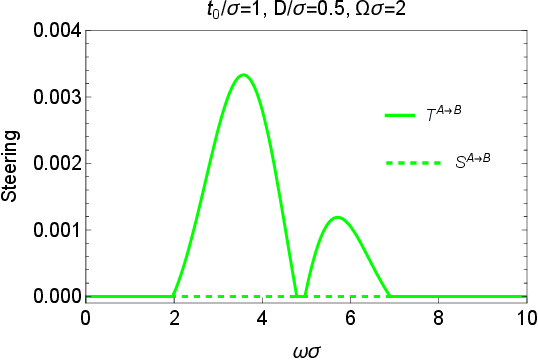}
\label{fig2d}
\end{minipage}%
\caption{Quantum steering of two detectors in the background of gravitational waves and in Minkowski spacetime as a function of the gravitational wave frequency $\omega\sigma$ for various values of the energy gap $\Omega\sigma$ and $t_{0}/\sigma$ with $D/\sigma=0.5$.}
\label{Fig.1}
\end{figure}

In Fig.\ref{Fig.1}, we plot quantum steering $T^{A\rightarrow B}$ (in the background of gravitational waves) and $S^{A\rightarrow B}$ (in Minkowski spacetime) as a function of the gravitational wave frequency $\omega\sigma$ for different energy gaps $\Omega\sigma$ and $t_{0}/\sigma$. We find that quantum steering $T^{A\rightarrow B}$ cannot be harvested from the vacuum at the lower gravitational wave frequency $\omega\sigma$, while quantum steering $S^{A\rightarrow B}$ can be harvested in Minkowski space. We also find that when the gravitational wave frequency $\omega\sigma$ is much greater than the energy gap $\Omega\sigma$, quantum  steerability $T^{A\rightarrow B}$ is approximately equal to quantum  steerability $S^{A\rightarrow B}$, which means that the gravitational wave does not influence the harvested steering significantly. From Fig.\ref{Fig.1}, we can see that when the frequency of gravitational waves is roughly equal to the energy gap of the detector, that is, when $\omega\approx \Omega$, a strong resonance effect will occur, leading to maximized harvest of quantum steering around this frequency. For $\Omega\sigma=2$, it is shown that quantum  steerability $S^{A\rightarrow B}$ equals zero,
while gravitational waves can create quantum steering harvested from the vacuum, reflecting that gravitational waves can broaden the parameter range for harvested steering. From Fig.\ref{Fig.1}, we also see that quantum  steerability $T^{A\rightarrow B}$ can be greater than or less than quantum  steerability $S^{A\rightarrow B}$, indicating that the gravitational wave can either amplify or degrade the harvested steerability. For a smaller energy gap $\Omega\sigma$,  quantum  steerability $T^{A\rightarrow B}$ for $t_{0}/\sigma=1$ is generally larger than quantum  steerability $T^{A\rightarrow B}$ for $t_{0}/\sigma=0$ in most parameter spaces. Conversely, for a larger energy gap $\Omega\sigma$, this situation is completely opposite to the above situation.
Additionally, for a larger gravitational wave frequency  $\omega\sigma$, Fig.\ref{Fig.1} shows that quantum steering  $T^{A\rightarrow B}$  is independent of  $t_{0}/\sigma$.

\begin{figure}[htbp]
\begin{minipage}[t]{0.5\linewidth}
\centering
\includegraphics[width=3.0in,height=5.2cm]{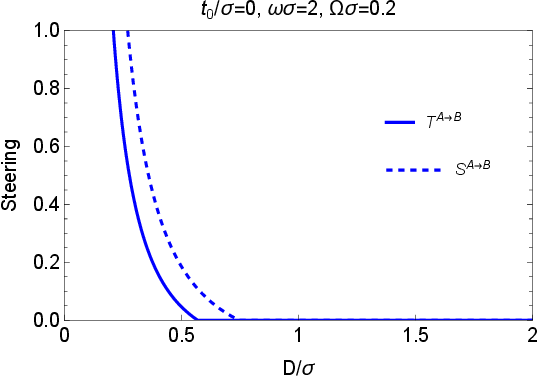}
\end{minipage}%
\begin{minipage}[t]{0.5\linewidth}
\centering
\includegraphics[width=3.0in,height=5.2cm]{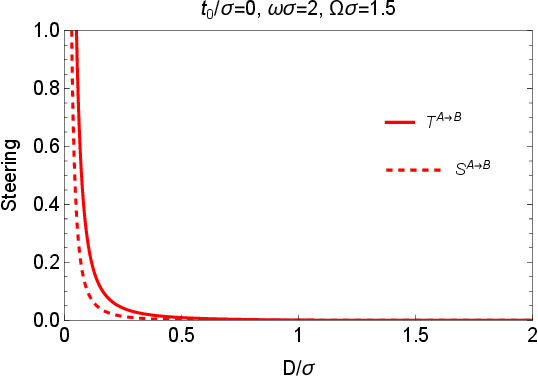}
\end{minipage}%

\begin{minipage}[t]{0.5\linewidth}
\centering
\includegraphics[width=3.0in,height=5.2cm]{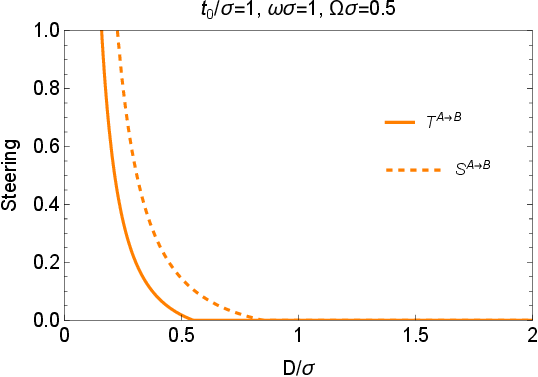}
\end{minipage}%
\begin{minipage}[t]{0.5\linewidth}
\centering
\includegraphics[width=3.0in,height=5.2cm]{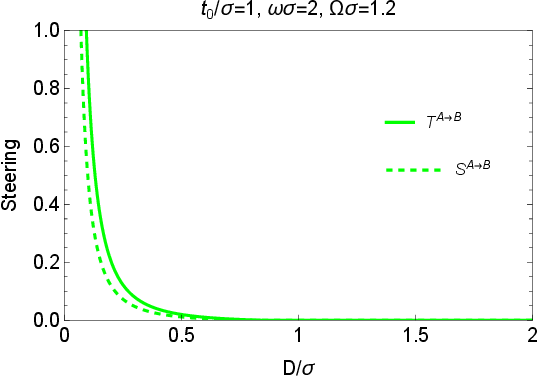}
\end{minipage}%
\caption{Quantum steering between two detectors in the background of gravitational waves and in Minkowski spacetime as a function of the detector separation $D/\sigma$.}\label{Fig.2(1)}
\end{figure}

In Fig.\ref{Fig.2(1)}, we illustrate the variation of quantum steering with respect to the detector separation $D/\sigma$. We find that harvested steering $T^{A\rightarrow B}$ and $S^{A\rightarrow B}$ first monotonically decrease and then suffer ``sudden death" with the increase of the detector separation $D/\sigma$, meaning that
reducing detector separation is beneficial for vacuum steering. From left column of Fig.\ref{Fig.2(1)}, we can see that the harvesting-achievable separation range of $T^{A\rightarrow B}$ is smaller than  the harvesting-achievable separation range of $S^{A\rightarrow B}$, indicating that  the presence of gravitational waves can shorten the harvesting-achievable separation range. From right column of Fig.\ref{Fig.2(1)}, we also see that  the harvesting-achievable separation range of $T^{A\rightarrow B}$ is wider than  the harvesting-achievable separation range of $S^{A\rightarrow B}$, showing that the presence of gravitational waves can  widen  the harvesting-achievable separation range.  From Fig.\ref{Fig.2(1)},  we can conclude that the presence of gravitational waves can broaden or narrow harvesting-achievable separation range of  vacuum steering, depending on  $t_{0}/\sigma$, $\Omega\sigma$,  and  $\omega\sigma$.

\begin{figure}
\begin{minipage}[t]{0.5\linewidth}
\centering
\includegraphics[width=3.0in,height=5.2cm]{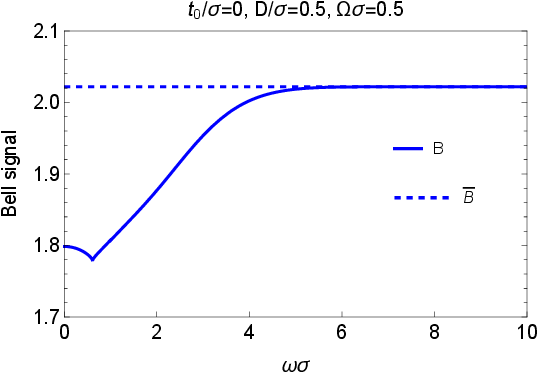}
\label{fig3a}
\end{minipage}%
\begin{minipage}[t]{0.5\linewidth}
\centering
\includegraphics[width=3.0in,height=5.2cm]{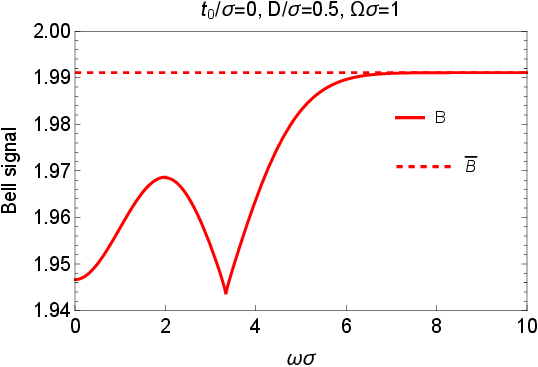}
\label{fig3b}
\end{minipage}%

\begin{minipage}[t]{0.5\linewidth}
\centering
\includegraphics[width=3.0in,height=5.2cm]{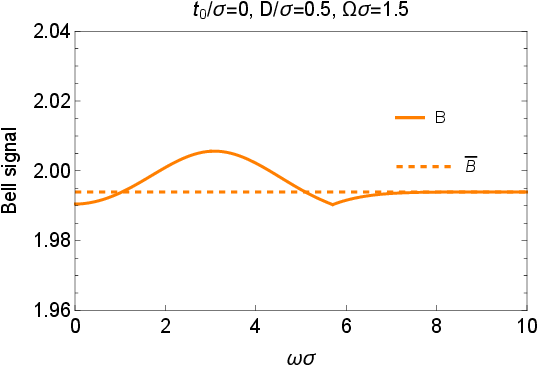}
\label{fig3c}
\end{minipage}%
\begin{minipage}[t]{0.5\linewidth}
\centering
\includegraphics[width=3.0in,height=5.2cm]{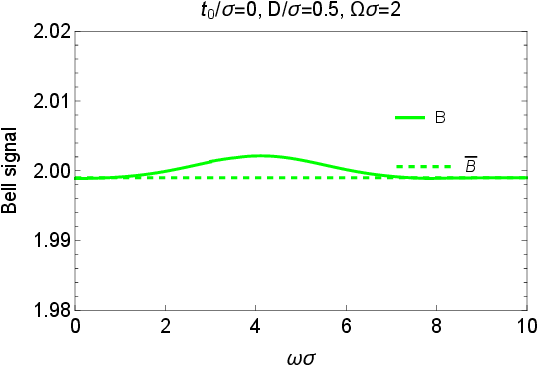}
\label{fig3d}
\end{minipage}%

\begin{minipage}[t]{0.5\linewidth}
\centering
\includegraphics[width=3.0in,height=5.2cm]{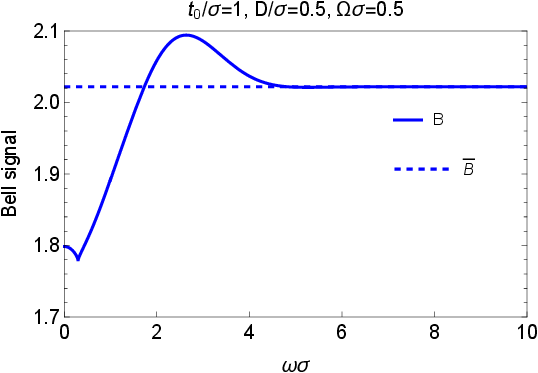}
\label{fig4a}
\end{minipage}%
\begin{minipage}[t]{0.5\linewidth}
\centering
\includegraphics[width=3.0in,height=5.2cm]{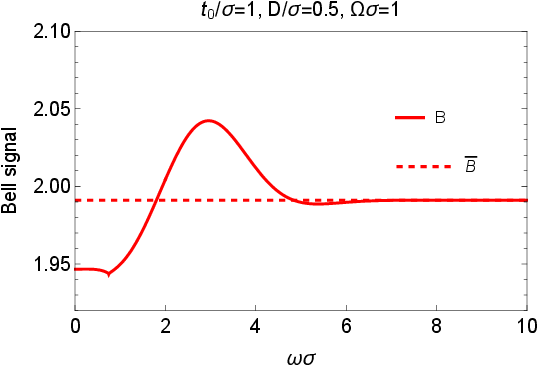}
\label{fig4b}
\end{minipage}%

\begin{minipage}[t]{0.5\linewidth}
\centering
\includegraphics[width=3.0in,height=5.2cm]{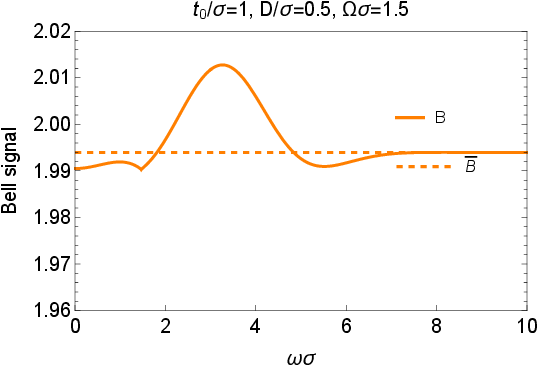}
\label{fig4c}
\end{minipage}%
\begin{minipage}[t]{0.5\linewidth}
\centering
\includegraphics[width=3.0in,height=5.2cm]{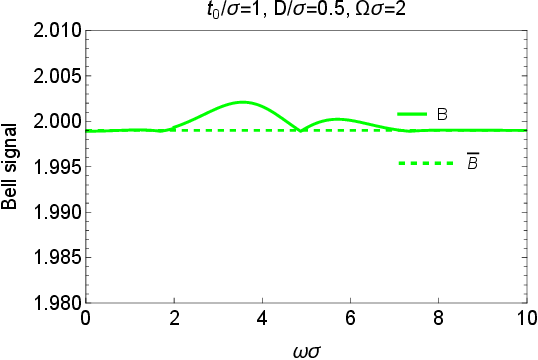}
\label{fig4d}
\end{minipage}%
\caption{Bell signal of two detectors under the influence of gravitational waves and in Minkowski spacetime is plotted as a function of the gravitational wave frequency $\omega\sigma$  for various values of the energy gap $\Omega\sigma$ and $t_{0}/\sigma$ with $D/\sigma=0.5$. }
\label{Fig.3}
\end{figure}

Similar to the research on vacuum steering, we also study Bell nonlocality for the bipartite system in the context of gravitational waves. Using Eqs.(\ref{S7}) and (\ref{AB}), the expression of Bell signal $B$ under the influence of gravitational waves can be found to be
\begin{equation}
B(\rho_{AB})=\max\{4\sqrt{2(|X|^2+|C|^2)}, {2\sqrt{4(|X|+|C|)^2+(|1-2P|-2|P|)^2}\}}.
\end{equation}
In order to facilitate the detection of the effects of gravitational waves on Bell nonlocality, we also explore Bell signal in Minkowski spacetime, labeled $\bar{B}$.

A more detailed study of the influence of gravitational waves on Bell signal  is shown in Fig.\ref{Fig.3}, where  $B$ and $\bar{B}$ are plotted as a function of the gravitational wave frequency $\omega\sigma$ for different $\Omega\sigma$ and $t_{0}/\sigma$.
From Fig.\ref{Fig.3}, we can see that Bell signal may be greater than 2 when $\omega\approx \Omega$. In other words, the system of two detectors that may violate the Bell inequality is considered to have Bell nonlocality when $\omega\approx \Omega$.
However, Bell signal is less than 2 regardless of the presence of the gravitational wave when $\Omega\sigma$ is not tuned to $\omega\sigma$. This means that the system of two detectors that does not violate the Bell inequality is considered to have no Bell nonlocality when $\Omega\sigma$ is not tuned to $\omega\sigma$. Therefore, we can conclude that, in most cases, the existence of gravitational waves does not produce Bell nonlocality harvested from the vacuum when $\Omega\sigma$ is not set to $\omega\sigma$. Consequently, harvested steering cannot be regarded as nonlocal.

\section{Conclusion}
The effect of gravitational waves on vacuum steering and Bell nonlocality harvested by two UDW detectors has been investigated. We obtain a description of how steering in the background of gravitational waves and in Minkowski spacetime vary with the gravitational wave frequency. It is shown that for smaller energy gaps, quantum steering cannot be harvested from the vacuum at the lower gravitational wave frequency, whereas it can be harvested in Minkowski spacetime.  However, for larger energy gaps,  quantum steering cannot be harvested from the vacuum in Minkowski spacetime, while the gravitational waves  can create net quantum steering.
This implies that gravitational waves can either amplify or diminish the harvested steerability, with the outcome depending on the energy gap. Quantum steering under the influence of gravitational waves is nearly equal to quantum steering in Minkowski spacetime when the gravitational wave frequency is much larger than the energy gap. The above phenomenon implies that the gravitational waves have no significant effect on steering harvesting. It is intriguing to find that a strong resonance effect for harvested steering occurs when the energy gap of the detectors is tuned to the frequency of the gravitational wave.

To explore the effect of gravitational waves on the harvesting-achievable range of vacuum steering,
we obtain a description of how steering changes with detector separation in the background of gravitational waves and in Minkowski spacetime. We find that the harvested steering first decreases monotonically and then experiences ``sudden death" as the detector separation increases.
We also find that the presence of gravitational waves can widen or narrow harvesting-achievable separation range of vacuum steering relative to Minkowski spacetime, which is related to the energy gap, the gravitational wave frequency, and the duration of the gravitational wave action. Furthermore,  the system of two detectors that may violate the Bell inequality is considered to exhibit Bell nonlocality when $\omega\approx \Omega$.
However, Bell signal is not more than 2, which means that  the system satisfies the Bell inequality when $\Omega\sigma$ is not tuned to $\omega\sigma$.  Therefore, harvesting steering cannot be regarded as nonlocal in
most parameter spaces.

\begin{acknowledgments}
This work is supported by the National Natural
Science Foundation of China (Grant Nos. 12205133), LJKQZ20222315, and Scientific and Technological Innovation Programs of Higher Education
Institutions of Shanxi Province, China (2023L327).
\end{acknowledgments}


\end{document}